\documentclass[10pt,a4paper]{article} 
\usepackage{lipsum}
\usepackage{url}
\usepackage[nochapters,subfig]{classicthesis} 
\usepackage[square,sort,comma,numbers]{natbib}
\usepackage{graphicx}
\graphicspath{{./}}
\usepackage{subfig}
\usepackage{booktabs}
\usepackage[smaller]{acronym}
\usepackage{amsmath}
\usepackage{nicefrac}
\usepackage{algorithm}
\usepackage{algorithmic}
\usepackage{textcomp}
\usepackage[T1]{fontenc}
\usepackage{xspace}
\newcommand*{\eg}{e.g.\@\xspace}
\newcommand*{\ie}{i.e.\@\xspace}
\newcommand{\FL}[1]{\textsc{fl#1}}
\definecolor{eham}{rgb}{1.0, 0.714, 0.467}
\definecolor{egll}{rgb}{0.427, 0.714, 1.0}
\definecolor{egss}{rgb}{0.859, 0.82, 0.0}
\definecolor{lgav}{rgb}{0.0, 0.427, 0.859}
\definecolor{lemd}{rgb}{0.573, 0.0, 0.0}
\definecolor{egkk}{rgb}{0.0, 0.286, 0.286}
\definecolor{lirf}{rgb}{0.141, 1.0, 0.141}
\definecolor{eddf}{rgb}{0.286, 0.0, 0.573}
\definecolor{eggw}{rgb}{0.714, 0.859, 1.0}
\definecolor{lfpg}{rgb}{0.573, 0.286, 0.0}

\newcommand{\airp}[1]{\textcolor{#1}{\textsc{#1}}}

\newcommand{\PELT}{\textsc{pelt}}
\newcommand{\DBSCAN}{\textsc{dbscan}}

\acrodef{FL}{Flight Level}
\acrodef{NM}{Nautical Miles}
\acrodef{TS}{Time Series}
\acrodef{ACC}{Area Control Centre}
\acrodef{ACF}{Autocorrelation function}
\acrodef{AIC}{Akaike Information Criterion}
\acrodef{ATM}{Air Traffic Management}
\acrodef{CCF}{Crosscorrelation function}
\acrodef{CDF}{Cumulative Distribution Function}
\acrodef{DDR}{Demand Data Repository}
\acrodef{ETA}{Expected Time of Arrival}
\acrodef{HDR}{High Density Rule}
\acrodef{MLE}{Maximum Likelihood Estimator}
\acrodef{IID}{Independent Identically Distributed}
\acrodef{PDF}{Probability Density Function}
\acrodef{PMF}{Probability Mass Function}
\acrodef{TCC}{Terminal Control Centre}
\acrodef{UTC}{Coordinated Universal Time}
\acrodef{ASMA}{Arrival Sequencing and Metering Area}
\acrodef{ECAC}{European Civil Aviation Conference}
\acrodef{ICAO}{International Civil Aviation Organization}
\acrodef{MBIC}{Modified Bayesian Information Criterion}
\acrodef{PSRA}{Pre-Scheduled Random Arrivals}

\usepackage{hyperref}

\begin{document}

    \pagestyle{plain}
    \title{\rmfamily\normalfont\spacedallcaps{Data-driven modelling and validation of aircraft inbound-stream at some major European airports}}
    \author{\spacedlowsmallcaps{Carlo Lancia}\footnote{Leiden University} \and \spacedlowsmallcaps{Guglielmo Lulli}\footnote{Lancaster University}}
    \date{\today} 

    \maketitle

    \begin{abstract}
        This paper presents an exhaustive study on the arrivals process at eight important European airports.
        Using inbound traffic data, we define, compare, and contrast a data-driven Poisson and \ac{PSRA} point process.
        Although, there is sufficient evidence that the interarrivals might follow an exponential distribution, this finding does not directly translate to evidence that the arrivals stream is Poisson.
        The main reason is that finite-capacity constraints impose a correlation structure to the arrivals stream, which a Poisson model cannot capture.
        We show the weaknesses and somehow the difficulties of using a Poisson process to model with good approximation the arrivals stream.
        On the other hand, our innovative non-parametric, data-driven \ac{PSRA} model, predicts quite well and captures important properties of the typical arrivals stream.
    \end{abstract}

    \section{Introduction}\label{sec:introduction}

    Air congestion is a regular and persistent phenomenon in the air traffic system in both the US and Europe.
    Over the years, air traffic demand increased at much faster pace with respect to the increment of air-traffic-system capacity.
    Although in the last decade we have witnessed a mitigation of congestion phenomena --air traffic demand has not revamped yet since the 2008 economic crisis-- the latest air traffic statistics published by \textsc{Eurocontrol} show a significant deterioration of on-time performance in the \acl{ECAC} area.
    The average delay per flight is at its highest since the last 10 years~\citep{coda2016}.
    To have a good grasp of the level of congestion, 7,167 flights were canceled and 107,426 were delayed in Europe between November 28 and December 27, 2016.
    The situation was even worse in the US, as the statistics of canceled and delayed flights showed double figures~(\citeauthor{flightstats}).
    However, for the sake of completeness, the number of controlled flights in the US is much larger: 15.3 million  in the US versus 9.9 million flights in Europe in 2015~\citep{EUCTRL-FAA2015}.

    Airports are the most relevant bottlenecks in the air traffic system.
    During periods of high demand, the \ac{ASMA} additional time --a proxy for the average arrival runway queuing time of the inbound traffic flow-- is  a good indicator indicacator of airport congestion. In 2015, the average \ac{ASMA} additional time at the top 30 European airports amounted to  2.27 minutes per arrival, increasing by about 18\% with respect to the previous year.
    The \ac{ASMA} performance deterioration in 2015 was largely driven by an increase in average additional \ac{ASMA} time at London Gatwick, Stockholm Arlanda, Dublin, and Brussels.
    London Heathrow has by far the highest level of average additional \ac{ASMA} time in Europe, which is almost 9 minutes per arrival, followed by London Gatwick with more than 4 minutes per flight~\citep{PRR2015}.
    Similar situations occur in the US, although with less contrast in additional time reported across airports~\citep{EUCTRL-FAA2015}.
    This situation occurs despite the fact that the principal airports in Western and Central Europe are treated as \emph{fully coordinated} meaning essentially that the number of flights that can be scheduled per hour (or other unit of time) is not allowed to exceed the \emph{declared capacity} of the airports~\citep{deNO2003}.
    In the U.S., scheduling limits are applied only to airports of the New York region, Washington Reagan, and Chicago O'Hare airport, under the \acl{HDR}.

    Airport congestion cannot be deemed as a recent phenomenon.
    Starting from the pioneering work of~\citet{Blum1959}, airport operations has attracted the interest of the scientific community in the attempt to alleviate congestion.
    Many quantitative methods have been developed to understand the various causes of congestion.
    These methods try to ameliorate the level of congestion by detecting possible actions for improving the use of capacity and reducing delays.
    In particular, a great amount of work has been devoted to study the arrivals process at airports and the corresponding queues.
    Given the nature of the phenomenon, most of these studies rely on either queuing theory or simulation models.
    As highlighted in~\citet{Koop1972}, models should include both
    \emph{i}) fluctuations of the arrivals demand over time due to hub-and-spoke operations carried out at major airports and
    \emph{ii}) randomness affecting the arrivals to provide good estimates of congestions.
    Indeed,~\citet{Koop1972} assumed that the statistics of arrivals follow a Poisson law, but with an arrival rate
    that is a strongly-varying function of time according to statistics actually observed at airports.
    As pointed out by~\citet{HO1975}, the assumption of Poisson arrivals for airport demand has two very appealing properties:
    \emph{i}) it is mathematically tractable and is consistent with observations at major airports, and
    \emph{ii}) it has been extensively used in the transportation literature.
    Indeed, this assumption has been adopted to study the arrival stream at several airports, e.g.,  J.F. Kennedy and La Guardia~\citep{Koop1972}, Toronto Person \citep{Bookbinder1986}, and Boston Logan~\citep{HO1975} among others; even though it  has been corroborated only in more recent times by  \citet{willemain2004statistical}.
    In that paper, the authors examine data on all arrivals to nine major US airports during December 2003 for evidence of exponentiality in the distribution of the \ac{ETA} as estimated when the aircraft were 100 miles from their destinations under the assumption that the arrival rates changes are relatively slow over time. The analysis confirmed the nearly-exponential character of the intervals between projected \acp{ETA}.

    However, as we are going to highlight in this paper, there are two inherent issues with Poisson arrivals. The first one is the inability of the Poisson process to capture any correlation between arrivals at consecutive time periods. This leads to an overestimation of the queue length~\citep{caccavale2014model}.
    Second, if we model the arrivals stream as a non-homogeneous Poisson process, a possibly large number of parameters has to be estimated, consistently with Koopman approach.
    In addition, the use of projected arrivals \citep{willemain2004statistical} might not be adequate in a European context, where the arrival rate at several major European airports tend to change \emph{rather fast}.

    To overcome these issues,~\citet{guadagni2011queueing} have recently proposed to model the arrival stream at airports with a Pre-Scheduled Random Arrivals (\ac{PSRA}) process, which is obtained from a deterministic schedule by superimposing \ac{IID} random delays.
    The list of actual arrival time is the result of ``mixing-up'' the fixed schedule by the addition of random perturbations.
    The resulting process --which was known since the 60's~\citep{Kendall1964}-- is able to provide very good fit with real data, as shown for London Heathrow airport by~\citet{caccavale2014model}.
    Moreover, the \ac{PSRA} process is easy to study numerically and some significant progress has been recently made on its mathematical properties~\citep{lancia2013advances}.
    \citet{nikoleris2012queueing} used the \ac{PSRA} to develop a single-server queuing model for trajectory-based aircraft operations that accounts explicitly for varying levels of imprecision in meeting prescribed times of arrival at airspace fixes.
    \citet{gwiggner2014data} have used and compared Poisson arrivals and \ac{PSRA} in a queuing analysis with the aim of describing the delays in Japanese traffic flow.

    In this paper, we present an exhaustive study on the arrivals process at major European airports, in the period from June 15 to September 15, 2016.
    In particular, we study the interarrival times on a large data set which includes some of the busiest and most congested airports in Europe, namely, London Heathrow (\ac{ICAO} code: \airp{egll}), London Gatwick (\airp{egkk}), Frankfurt am Main (\airp{eddf}), Amsterdam Schiphol (\airp{eham}), and Paris C. De Gaulle (\airp{lfpg}).
    As we are also interested in the modeling of medium-intensity traffic, we include in the dataset arrivals at three other important airports: Madrid Barajas (\airp{lemd}), Rome Fiumicino (\airp{lirf}), and Athens International (\airp{lgav}).

    Using data on arrivals at the airports listed above, we compare and contrast the Poisson process with \ac{PSRA}.
    The interarrival data seemingly suggest that the underlying arrivals stream might be Poisson.
    Nevertheless,  using such a process to model the arrivals stream with good approximation presents some weaknesses and difficulties, which we describe in detail in the following sections.
    On the other hand, the \ac{PSRA} combines good predictive qualities of the arrivals stream at airports with the desirable property of being non-parametric or low-dimensional.

    The contribution of this paper is three-fold:
    \emph{i}) it validates the results of~\cite{willemain2004statistical} in a European context;
    \emph{ii}) it proposes an innovative data-driven approach to the modeling of the inbound stream, showing all procedural details to define a non-homogeneous Poisson process and a \ac{PSRA};
    \emph{iii}) it compares the processes obtained this way on a solid statistical ground.

    The remainder of this paper is organized as follows.
    In Section~\ref{sec:dataset_description}, we describe the dataset and the data-analysis methodology used for this study.
    Section~\ref{sec:results} presents the finding of the paper:
    in Sections~\ref{sec:exp} and~\ref{sec:serial_corr}, we present a thorough statistical analysis on the arrivals stream at the considered airports; in Section~\ref{sec:data_driven_modelling_of_the_inbound_flow}, we show how to construct a time-dependent Poisson process in a data-driven manner; in Section~\ref{sub:data_driven_psra}, we compare and contrast it with our innovative data-driven \ac{PSRA} approach.
    Finally, in Section~\ref{sec:conclusions} we discuss the results and provide closing comments.

    \section{Dataset description}\label{sec:dataset_description}

    Inbound flights data were extracted from \textsc{Eurocontrol}'s \ac{DDR} between June 15 and September 15, 2016.
    The choice of the summer period aims at reducing the variability of traffic patterns, which might be caused by adverse weather conditions.
    Table~\ref{tab:flights_count} displays the total count of inbound flights in the study sample for each airport.

    \begin{table}[tbp]
      \centering
      \begin{tabular}{lcr}
\toprule
{} & \acs{ICAO} code &  sample size \\
airport name                            &                 &              \\
\midrule
Frankfurt am Main International Airport &     \airp{eddf} &        58167 \\
London Gatwick Airport                  &     \airp{egkk} &        39746 \\
London Heathrow Airport                 &     \airp{egll} &        56716 \\
Amsterdam Airport Schiphol              &     \airp{eham} &        63279 \\
Madrid Barajas International Airport    &     \airp{lemd} &        48162 \\
Charles de Gaulle International Airport &     \airp{lfpg} &        60122 \\
Athens International Airport            &     \airp{lgav} &        29503 \\
Rome Fiumicino International Airport    &     \airp{lirf} &        43333 \\
\bottomrule
\end{tabular}

      \caption{Size of study sample for each airport considered. Observation period: from June 15 to September 15, 2014.}\label{tab:flights_count}
    \end{table}

    We query the \ac{DDR} to extract both \emph{M1 and M3 flight plans}, respectively the last flight plan agreed with \textsc{Network Manager} (\textsc{Eurocontrol}) and the flight trajectory actually flown.
    From M1 and M3 flight plans, we obtain \(t^{M1}\) and \(t^{M3}\), the time at which the aircraft enter a cylinder of 40 NM (Nautical Mile) around the airport.
    Unless explicitly stated, times are local.

    We consider the passage time at 40 NM as a proxy for the time when the flight starts the approach phase and is handed over to the Terminal Control.
    This time could have been measured with more accuracy by considering the instantaneous latitude and longitude of each aircraft as reported by the \ac{DDR}.
    However, this approach to data extraction would have depended on the specific airport under consideration --an added complication that does not match with the aim of this paper.
    Regardless of the data-querying method adopted, the general data-analysis methodology illustrated hereafter remains in fact valid.

    The timestamps of the passage at 40 NM form a \ac{TS} for each airport.
    Once the \ac{TS} has been created, \emph{interarrival times} are defined as the time lapse in seconds between two successive events. As the arrivals rate is not constant, this \ac{TS} has no fixed frequency.
    Interarrivals are measured as precisely as the nearest second, and ties are present in the data, \ie{} a set of two (or more) equal values.

    We investigate evidence of exponentiality in the interarrivals by plotting empirical quantiles against the theoretical quantiles of a Weibull distribution
    \begin{equation}
        f_W(x; \lambda, \beta) = \frac{\beta}{\lambda} {\left(\frac{x}{\lambda}\right)}^{\beta-1} e^{{-(\frac{x}{\lambda})}^\beta} \,,
        \label{eq:WeibullPDF}
    \end{equation}
    as proposed by~\citet{willemain2004statistical}.
    The use of a Weibull is preferable over an exponential because the \emph{shape parameter} \(k\) can appreciably modify the probability of observing small interarrivals, \ie{} large number of arrivals in a fixed interval, while the chance of observing large interarrivals still decays exponentially fast.
    The presence of ties in the sample is overcome by using the discrete version of~\eqref{eq:WeibullPDF}~\citep{nakagawa1975discrete,barbiero2013discrete}
    \begin{equation}
        P_W(X=x;q,\beta) = q^{x^\beta} - q^{{(x+1)}^\beta},
        \label{eq:DWeibullPMF}
    \end{equation}
    where the parameter \(\lambda\) corresponds to \(-\log^{-\nicefrac{1}{\beta}}q\).
    The special cases of an exponential distribution and a geometric \ac{PMF} are obtained respectively from~\eqref{eq:WeibullPDF} and~\eqref{eq:DWeibullPMF} by setting \(\beta = 1\).
    QQ-plots are drawn for three different time intervals, namely, 08:00--09:30, 12:00--13:30, and 18:00--19:30, local time; these are meant to capture different operational phases of the airports considered in this study, especially for those hosting hub-and-spoke operations.
   The Kolmogorov-Smirnov test~\citep{taylor2011nonparam} is used to test for goodness of fit.

    Typical characteristics of the inbound stream are explored by aggregating the arrivals \ac{TS} by intervals of ten minutes.
    Fluctuations in the daily evolution of the mean demand were described by 99\% pointwise confidence intervals.
    We look for evidence of serial correlations in the arrivals stream by computing the \ac{ACF} on the premise that the capacity of both en-route sectors (airspace) and airports impose constraints on the number of arrivals in consecutive time intervals.
    Stationarity of the arrivals \acs{TS} is achieved by taking first-order differences and then checked in a 24-hour window by the augmented Dickey-Fuller test~\citep{fuller2009introduction,seabold2010statsmodels}.

    As mentioned in \S\ref{sec:introduction}, we compare and contrast two classes of models for the inbound stream: a non-homogeneous Poisson process and a \ac{PSRA}.
    However, instead of modeling the Poisson intensity \(\lambda(t)\) or the distribution of the \ac{PSRA} delays, we adopt the following data-driven modeling procedures.
    We approximate the intensity of the time-dependent Poisson process with a step-function. The intensities and the corresponding time intervals are computed using both the \PELT{} algorithm~\citep{killick2012optimal}
    --to detect changepoints in the arrivals stream, \ie{} the time points where either the mean or the variance of the arrivals \ac{TS} undergoes a \emph{structural} change--
    and \DBSCAN{}~\citep{ester1996density,pedregosa2011scikit} to cluster  the changepoints and the estimated intensities in the \((t, \lambda)\) plane.
    Because the behavior and performance of \PELT{} strongly depend on the settings of the penalty function, we compared both the penalties based on \ac{AIC}~\citep{akaike1973information} and \ac{MBIC}~\citep{zhang2007modified}. Our feeling is that the \ac{AIC} penalty tends to produce a quite large Type II error while \ac{MBIC} tends to give a large Type I error.
    In our statistical analysis, we decided to use penalties based on \ac{AIC}.
    A comparison of the sensitivity of the \ac{AIC} penalty versus the default \ac{MBIC}~\citep{zhang2007modified} is provided in the Appendix.
    The details of the procedure are described in the listing below.

    \begin{algorithm}
        \begin{algorithmic}[1]
            \STATE \(B \leftarrow \) time series of binned arrivals
            \STATE \(H_0 \leftarrow \) `Poisson'
            \STATE \(T, L \leftarrow \) PELT\((B, H_0)\)
            \STATE \(n, C \leftarrow \) DBSCAN\((T, L)\)
            \STATE \COMMENT{\(C\) is a list of length \(n\) that contains 2-dimensional lists of clustered \(t, \lambda\)-points}
            \STATE \(\hat{T} \leftarrow [~] \)
            \STATE \(\hat{L} \leftarrow [~] \)
            \FOR { \(i = 1\) \TO \(n\) }
                \STATE \(\hat{t}_k, \hat{l}_k \leftarrow \) \texttt{centroid(C[i])}
                \STATE Append \(\hat{t}_k\) to \(\hat{T}\)
                \STATE Append \(\hat{l}_k\) to \(\hat{L}\)
            \ENDFOR
            \STATE \(\lambda^\ast(t) \leftarrow \) step\((\hat{T}, \hat{L})\)
            \COMMENT{A step function taking on value \(\hat{\lambda}_i\) for \(\hat{t}_i \leq t < \hat{t}_{i+1}\)}
            \RETURN \(\lambda^\ast(t)\)
        \end{algorithmic}
        \caption{Data-driven \emph{Poisson} process.}\label{Alg:POISSON}
    \end{algorithm}

    \noindent
    For the \ac{PSRA}, we simulate the process
    \begin{equation}
        \label{eq:psra-like}
        t_i = t^{M1}_i + \xi_i \,,
    \end{equation}
    where
    \begin{itemize}
        \item \(\xi_i\) is extracted \emph{with replacement} from the empirical distribution of the delays \({\{t^{M3}_i - t^{M1}_i\}}_i\);
        \item \(t^{M3}_i\) is the \emph{effective} arrival time of the \(i\)th aircraft at 40 NM\@;
        \item \(t^{M1}_i\) is its \emph{intended} arrival time at 40 NM\@.
    \end{itemize}
    The simulation procedure is detailed in the following:

    \begin{algorithm}
        \begin{algorithmic}[1]
            \STATE \(T^{M1} \leftarrow \) sequence of \emph{intended} M1 arrival times (\(t_i^{M1}\))
            \STATE \(T^{M3} \leftarrow \) sequence of \emph{estimated} M3 arrival times (\(t_i^{M3}\))
            \STATE \(\Xi \leftarrow T^{M3} - T^{M3}\)
            \STATE \(T \leftarrow [~] \)
            \FOR {\(t\) in \(T^{M1}\)}
                \STATE \(\xi \leftarrow \) random sample extracted from \(\Xi\)
                \STATE Append \(t + \xi\) to \(T\)
            \ENDFOR
            \RETURN \(T\)
        \end{algorithmic}
        \caption{Data-driven \acs{PSRA} process.}\label{Alg:PSRA}
    \end{algorithm}

    The two models above are compared with respect to their capability of reproducing the average daily arrivals and the serial correlations of the inbound stream.
    The latter are measured by the Pearson product-moment correlation coefficient \(\rho\) for the number of arrivals at 40 NM in consecutive intervals of 10 minutes, \ie{}
    \begin{equation}
        \label{eq:pearson_rho}
        \rho_{t_i, t_{i+1}} = \frac{\text{cov}(N_i, N_{i+1})}{\sigma(N_i) \, \sigma(N_{i+1})} \,,
    \end{equation}
    where \(|t_{i+1} - t_i| = 10 \text{min}\), \(N_{i}\) is the number of arrivals at 40 NM between \(t_{i}\) and \(t_{i+1}\), and \(\text{cov}(\cdot, \cdot)\) denotes covariance while \(\sigma(\cdot)\) denotes standard deviation.
    Covariance and standard deviations in~\eqref{eq:pearson_rho} are computed with respect to the longitudinal axis of the dataset, \ie{} over each day included in the study period.

    All figures and statistical analyses were produced using \texttt{Python v.3.6} and \texttt{R v.3.3} (via \texttt{rpy2}).

    \section{Results}\label{sec:results}

        \subsection{Exponentiality of the interarrival times}\label{sec:exp}

\begin{figure}
            \centering
            \subfloat[Arrivals at \FL{240} in the time period 08:00--09:30 local time.]{
                \includegraphics[width=.9\textwidth]{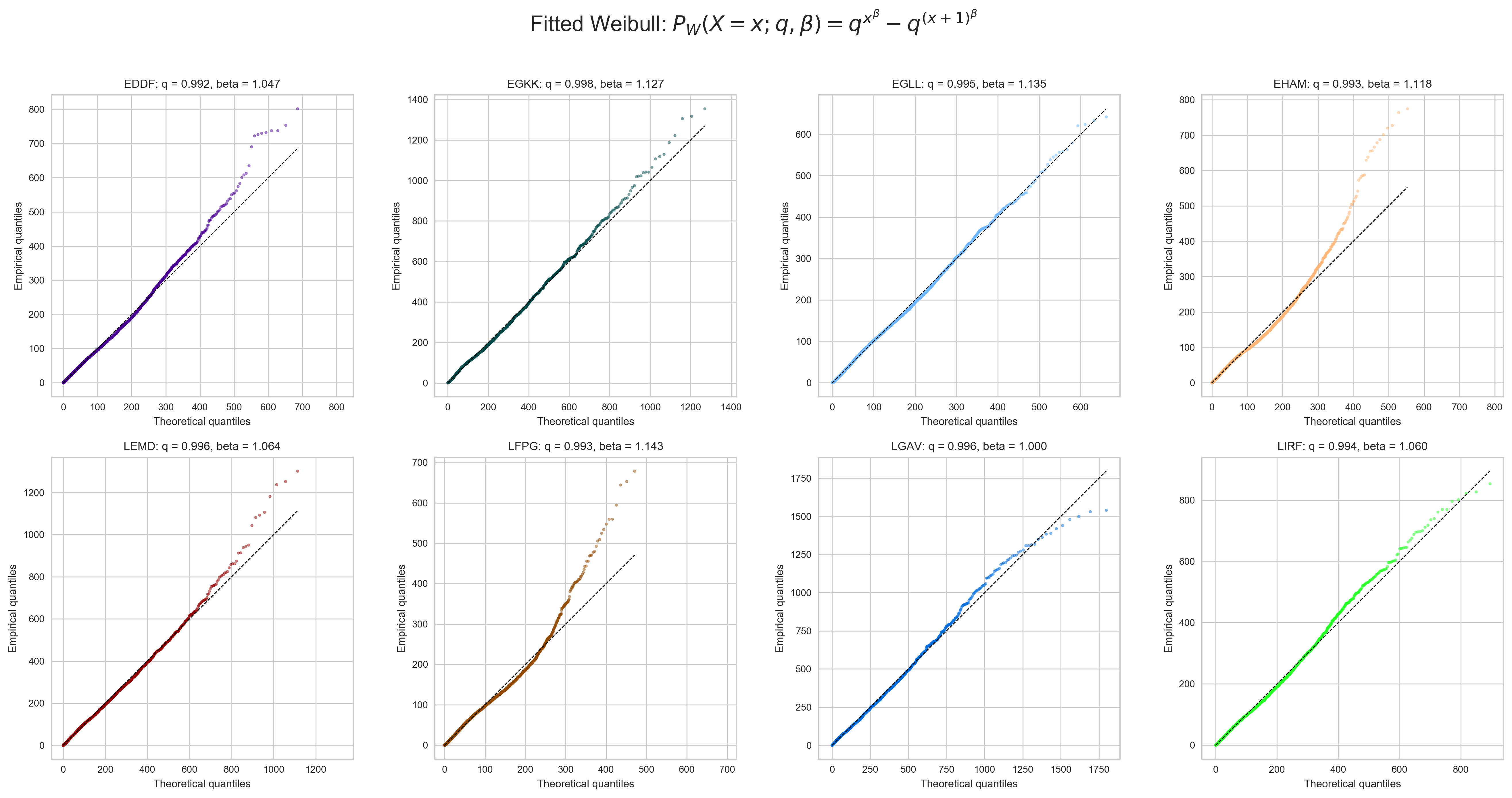}\label{fig:qqplot5-8}
            }\\
            \subfloat[Arrivals at \FL{240} in the time period 12:00--13:30 local time.]{
                \includegraphics[width=.9\textwidth]{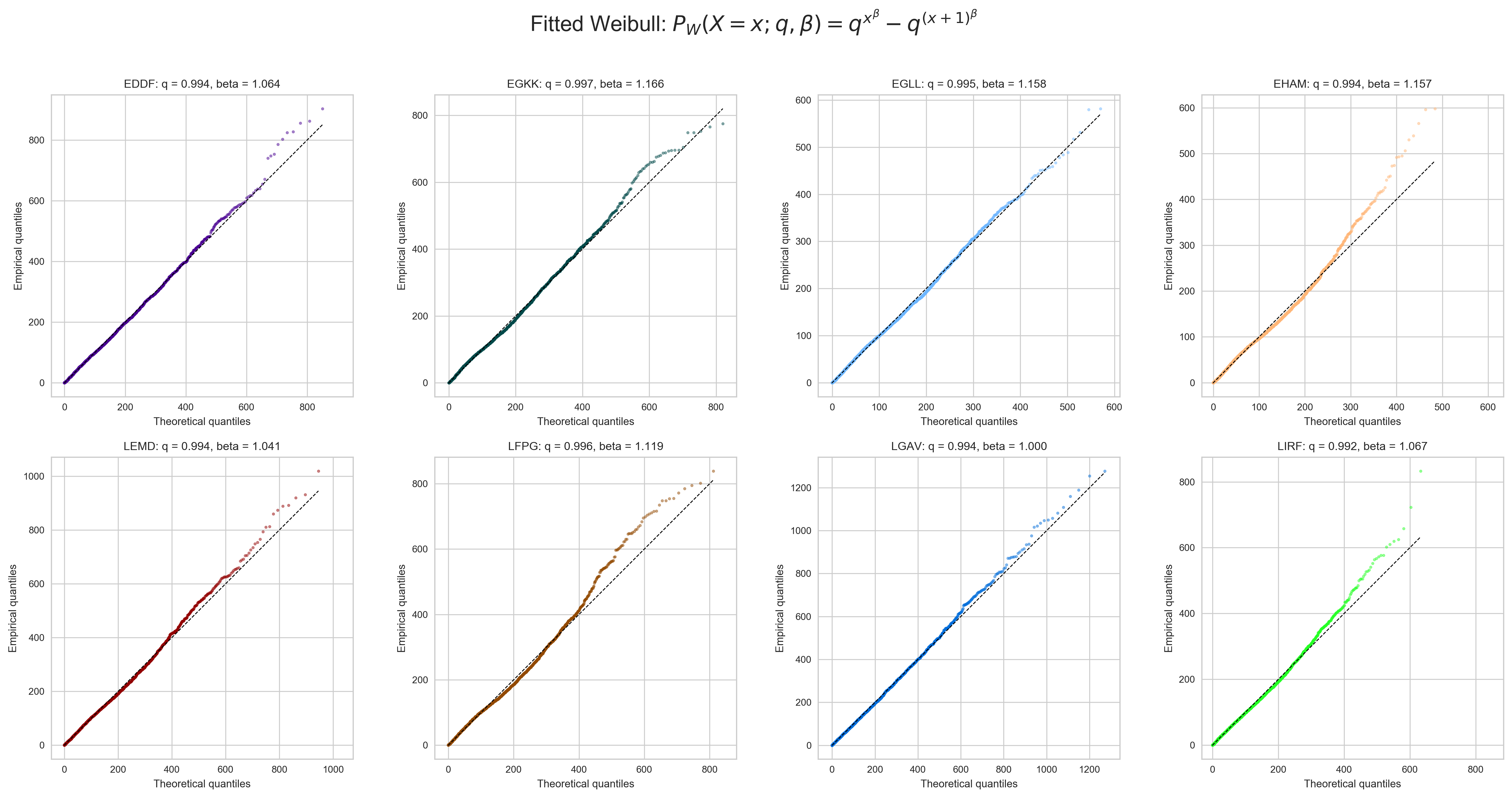}\label{fig:qqplot5-11}
            }\\
            \subfloat[Arrivals at \FL{240} in the time period 18:00--19:30 local time.]{
                \includegraphics[width=.9\textwidth]{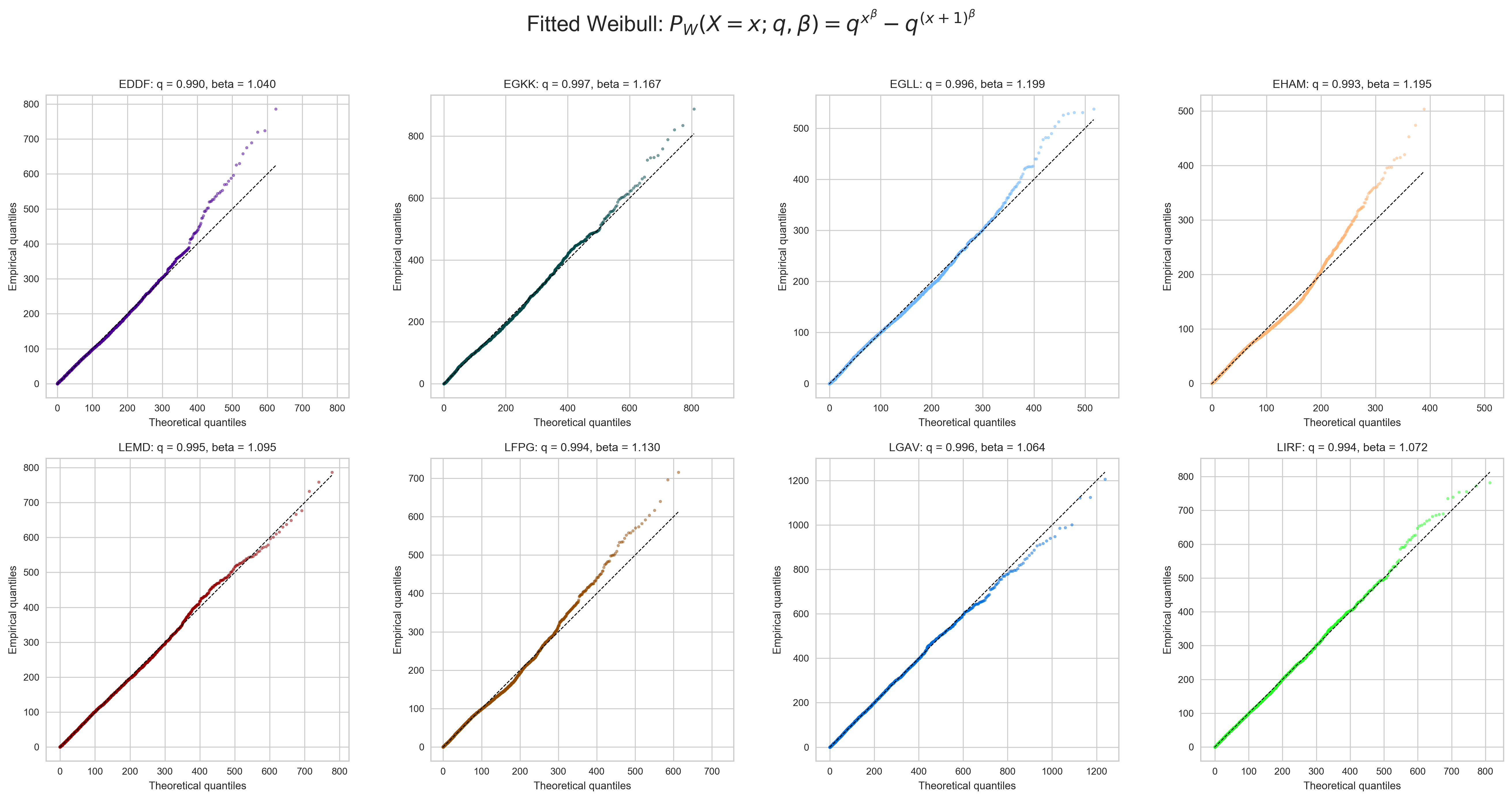}\label{fig:qqplot5-17}
            }
            \caption{Plot of empirical quantiles of interarrivals at \FL{240} versus theoretical quantiles from a discrete Weibull \ac{PMF}~\eqref{eq:DWeibullPMF}, whose parameters are listed in Table~\ref{tab:fitted}.}\label{fig:qqplot}
        \end{figure}

        Figure~\ref{fig:qqplot} shows, for each of the eight airports and three time intervals considered, the QQ-plot of the interarrivals against the corresponding fitted Weibull~\eqref{eq:DWeibullPMF}.
        Regardless of the time interval, there is quite a good accordance between empirical and theoretical quantiles in the bulk of the distribution.
        This can be observed as a general flat adherence of the QQ-plot onto the 45-degrees dotted line and should be interpreted as the capability of Weibull interarrivals to describe small-to-moderate interarrival times, \ie{} situations of high demand.
        However, the goodness of the fit deteriorates on the tails and it typically shows overdispersion, which can be severe at \airp{eddf}, \airp{eham}, \airp{lemd}, and \airp{lfpg}.
        A remarkable exception is \airp{egll}, for which the demand fluctuates around the value of 40 aircraft/hour (corresponding to an average interarrival of 90 seconds).
        Accordingly, \airp{egll} exhibits the smallest degree of overdispersion on the tails among the airports considered in this paper.

        Table~\ref{tab:fitted} reports the parameters \(q\) and \(\beta\), the mean of the fitted distribution, the Kolmogorov-Smirnov \(D\)-statistic, and the \(p\)-value of the corresponding goodness-of-fit test for each time interval and airport considered.
        The fitted shape-parameter \(\beta\) is always fairly close to 1, meaning that the fitting Weibull looks like an exponential/geometrical.
        Very often we can reject the null hypothesis of Weibull interarrivals at the 1\% significance level.
        This is due to the large size of the sample considered, which makes the test very powerful.

        \begin{table}
            \centering
            \begin{tabular}{llrrrrr}
\toprule
             &             &   $q$ & $\beta$ &    mean & $D$-stat. & $p$-value \\
time & airport &       &         &         &           &           \\
\midrule
08:00--09:30 & \airp{eddf} & 0.992 &   1.046 &  93.131 &     0.020 &      0.03 \\
             & \airp{egkk} & 0.998 &   1.127 & 214.506 &     0.032 &      0.02 \\
             & \airp{egll} & 0.995 &   1.135 & 103.273 &     0.024 &     <0.01 \\
             & \airp{eham} & 0.993 &   1.118 &  82.209 &     0.033 &     <0.01 \\
             & \airp{lemd} & 0.996 &   1.064 & 166.540 &     0.019 &      0.21 \\
             & \airp{lfpg} & 0.993 &   1.143 &  71.336 &     0.033 &     <0.01 \\
             & \airp{lgav} & 0.996 &   1.000 & 261.533 &     0.021 &      0.40 \\
             & \airp{lirf} & 0.994 &   1.060 & 129.793 &     0.023 &      0.03 \\
12:00--13:30 & \airp{eddf} & 0.994 &   1.064 & 122.668 &     0.018 &      0.13 \\
             & \airp{egkk} & 0.997 &   1.166 & 137.588 &     0.033 &     <0.01 \\
             & \airp{egll} & 0.995 &   1.158 &  90.262 &     0.026 &     <0.01 \\
             & \airp{eham} & 0.994 &   1.157 &  74.990 &     0.035 &     <0.01 \\
             & \airp{lemd} & 0.994 &   1.041 & 133.073 &     0.016 &      0.29 \\
             & \airp{lfpg} & 0.996 &   1.119 & 127.154 &     0.027 &     <0.01 \\
             & \airp{lgav} & 0.994 &   1.000 & 175.291 &     0.013 &      0.74 \\
             & \airp{lirf} & 0.992 &   1.067 &  88.120 &     0.021 &      0.01 \\
18:00--19:30 & \airp{eddf} & 0.990 &   1.040 &  82.466 &     0.015 &      0.12 \\
             & \airp{egkk} & 0.997 &   1.167 & 135.505 &     0.034 &     <0.01 \\
             & \airp{egll} & 0.996 &   1.199 &  85.679 &     0.031 &     <0.01 \\
             & \airp{eham} & 0.993 &   1.195 &  61.895 &     0.033 &     <0.01 \\
             & \airp{lemd} & 0.995 &   1.095 & 116.381 &     0.013 &      0.43 \\
             & \airp{lfpg} & 0.995 &   1.130 &  93.981 &     0.027 &     <0.01 \\
             & \airp{lgav} & 0.996 &   1.064 & 189.090 &     0.013 &      0.77 \\
             & \airp{lirf} & 0.994 &   1.072 & 117.987 &     0.016 &      0.21 \\
\bottomrule
\end{tabular}

            \caption{Parameters and goodness of fit of the fitted discrete Weibull \ac{PMF}~\eqref{eq:DWeibullPMF}. The parameter \(\beta\) is measured in seconds. The interpretation of the \(p\)-values of the Kolmogorov-Smirnov goodness-of-fit test is not straightforward due to the big sample size.}\label{tab:fitted}
        \end{table}

        \subsection{Properties of the inbound stream}\label{sec:serial_corr}

\begin{figure}
            \centering
            \includegraphics[width=\textwidth]{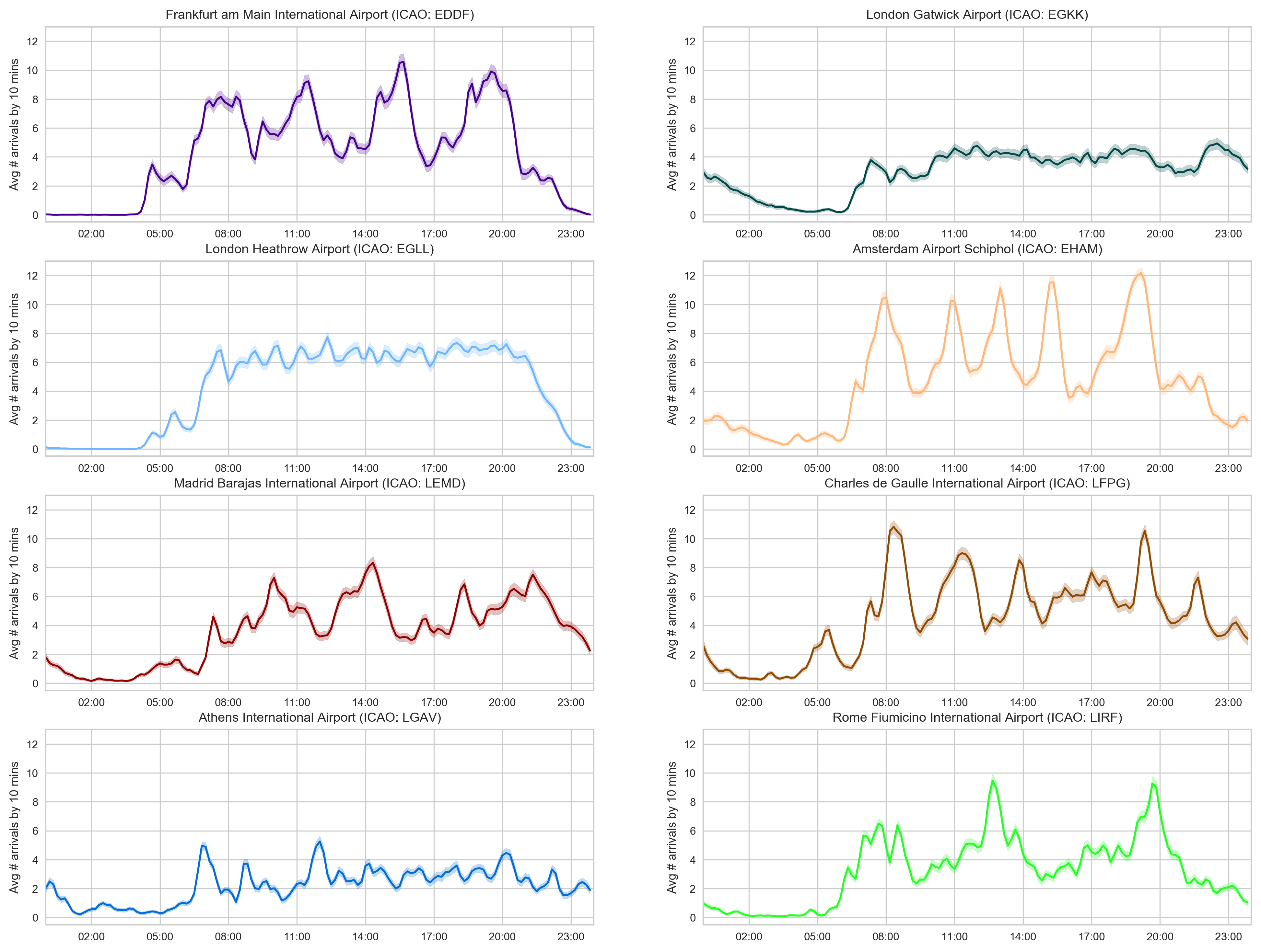}
            \caption{Average demand aggregated by 10 min. Shaded areas display 99\% point-wise confidence bands.}\label{fig:AvgArrivals}
        \end{figure}

        Figure~\ref{fig:AvgArrivals} shows the average demand over the day when aggregated by intervals of 10 minutes.
        The figure shows the mean number of arrivals along with 99\% pointwise confindence intervals.
        For many of the airports considered, the plot of the average daily arrivals shows the characteristic \emph{wavy pattern} typical of airports hosting hub-an-spoke operations.
        A notable exception is Heathrow, which has a nearly constant arrival rate and thus shows the best fit with a nearly-exponential distribution in Figures~\ref{fig:qqplot5-8}--\ref{fig:qqplot5-17}.

\begin{figure}
            \includegraphics[width=\textwidth]{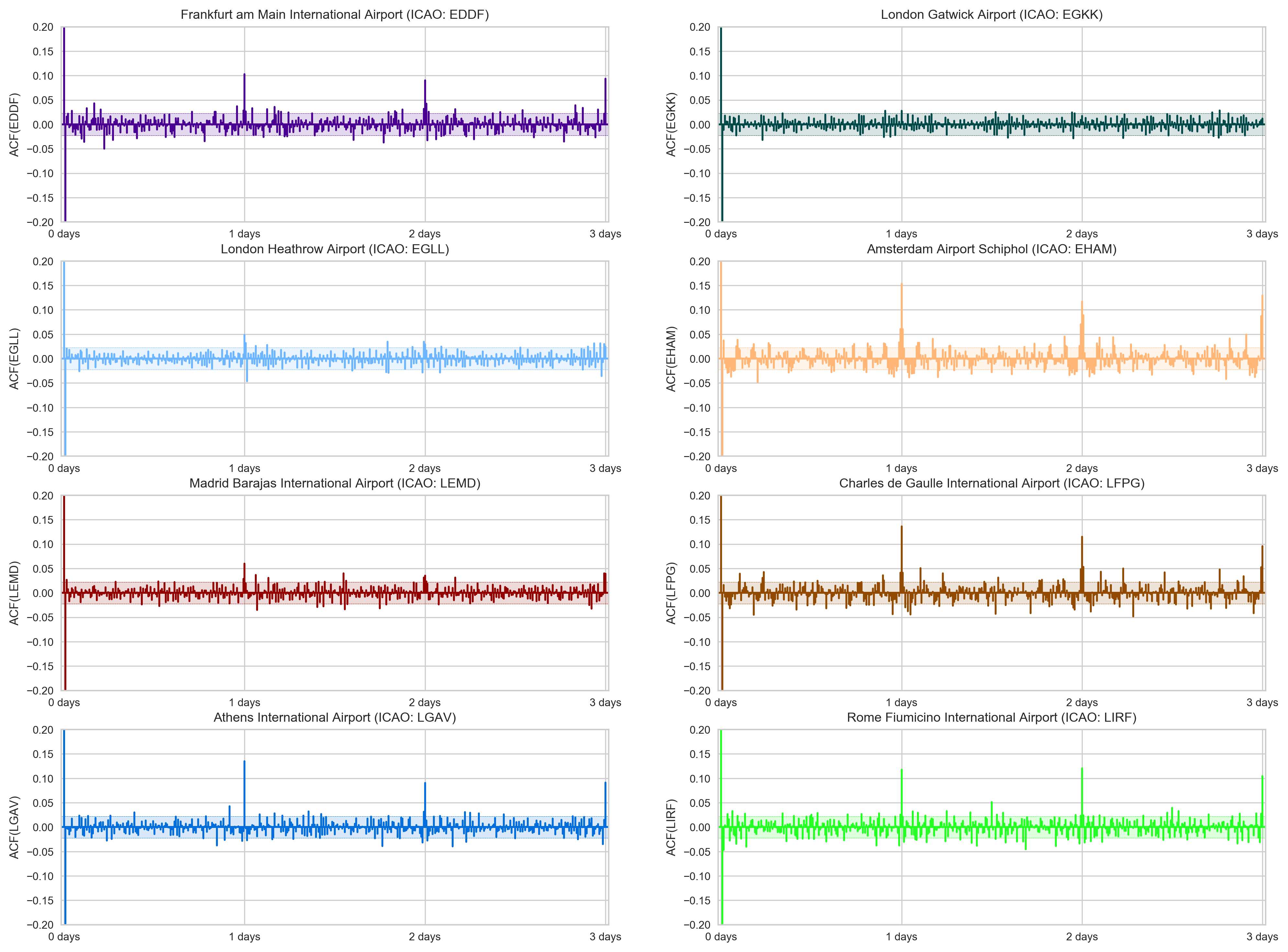}
            \caption{Autocorrelation of arrivals \ac{TS} aggregated by intervals of 10 minutes.
            Shared areas mark the limits of statistical significance at the 99\% level by Bartlett's Formula. For the sake of readability, the \(y\)-axis of the autocorrelation (1st column) is capped at \(\pm 0.2\). The values of the lag-0 and lag-1 correlations are off scale and reported by Table~\ref{tab:lag01}.}\label{fig:autocorr}
        \end{figure}

        The correlograms in Figure~\ref{fig:autocorr} highlight the presence of two kinds of serial correlation in the demand.
        All airports present a statistically significant strong negative lag-1 correlation.
        These correlations cannot be appreciated in full from Figure~\ref{fig:autocorr} --the \(y\)-axis is clipped at \(\pm 0.2\) for making other correlations more readable-- and are reported by Table~\ref{tab:lag01}.
        The presence of these correlations suggests that the net variation of the demand over an interval of 10 minutes is negatively correlated with the demand variation in the following 10 minutes.
        In other words, intervals where the demand increases (resp.\ decreases) are more likely to be followed by intervals where the demand decreases (resp.\ increases).
        This property can be also observed in Figure~\ref{fig:AvgArrivals} and demonstrates that the demand is bounded by capacity.
        In fact, if intervals of increased demand were likely to be followed by another interval of increased demand, then the capacity of the airport could be temporarily exceeded.

\begin{figure}
            \includegraphics[width=\textwidth]{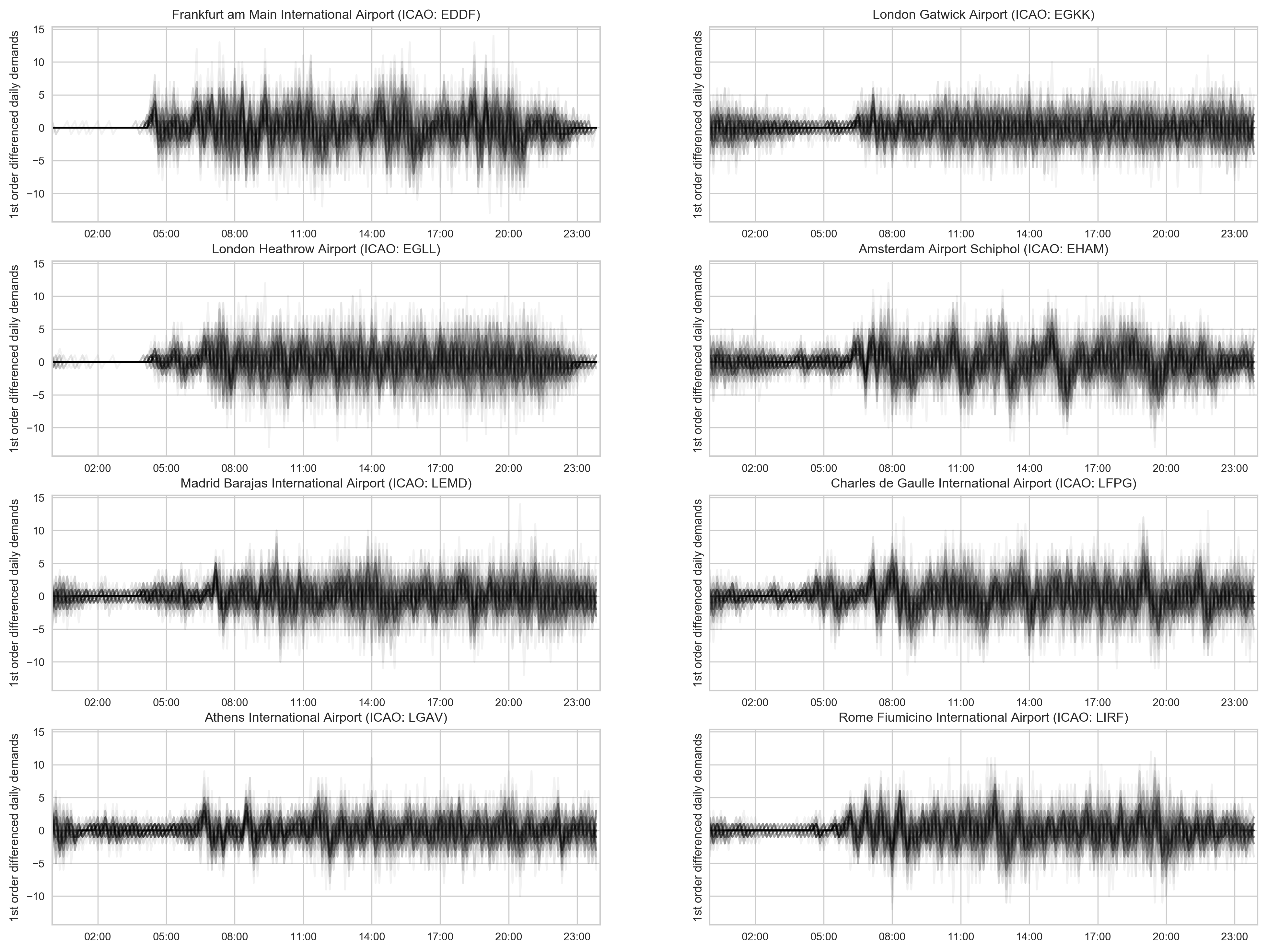}
            \caption{Differenced \acp{TS} of the daily demand. Each line corresponds to the demand of a day in the period under consideration. The demand in first-order differenced to obtain stationarity. Lines are drawn with a high transparency, so that the opacity of the figure give a sense of the daily-periodicity strength.}\label{fig:diff1demand}
        \end{figure}

        Many airports show the presence of statistically significant correlations at lags of 1, 2, and 3 days.
        These correlations are not strong in absolute terms, yet they are the strongest shown by the correlograms, and their relatively low magnitude can be explained by the presence of natural daily variation of the demand evolution in a quite large sample.
        Further, graphical inspection of daily-demand data seems to suggest a cyclic evolution of the demand over days.
        Figure~\ref{fig:diff1demand} shows first-order differenced \acp{TS} of the daily demand.
        Lines are drawn with a high transparency, so that the opacity of the figure give a sense of the daily-periodicity strength.
        The figure indicates that it makes sense to consider of practical significance correlation at the lag of 1 day.
        Significant correlations can be find at other lags but they do not seem to follow any fixed pattern.
        In this respect, they are specific of the airport considered and, most likely, they are linked to the wavy evolution of the demand highlighted above.
        Evidence of 24-hour periodicity is weaker at \airp{egkk}, \airp{egll}, and \airp{lemd}.
        Indeed, for these airports, the differenced daily demand  looks like the superimposition of white noise signals, see Figure~\ref{fig:diff1demand}.


        \begin{table}[tbp]
            \centering
            \caption{Values of lag-0 and lag-1 autocorrelations.}\label{tab:lag01}
            \begin{tabular}{lr}
            \toprule
            \acs{ICAO} code & lag-1 autocorr.\\
            \midrule
            \airp{eddf} & -0.447\\
            \airp{egkk} & -0.526\\
            \airp{egll} & -0.440\\
            \airp{eham} & -0.359\\
            \airp{lemd} & -0.466\\
            \airp{lfpg} & -0.415\\
            \airp{lgav} & -0.479\\
            \airp{lirf} & -0.535\\
            \bottomrule
            \end{tabular}
        \end{table}

        \subsection{Data-driven Poisson process for the inbound air traffic flow}\label{sec:data_driven_modelling_of_the_inbound_flow}

\begin{figure}
            \includegraphics[width=\textwidth]{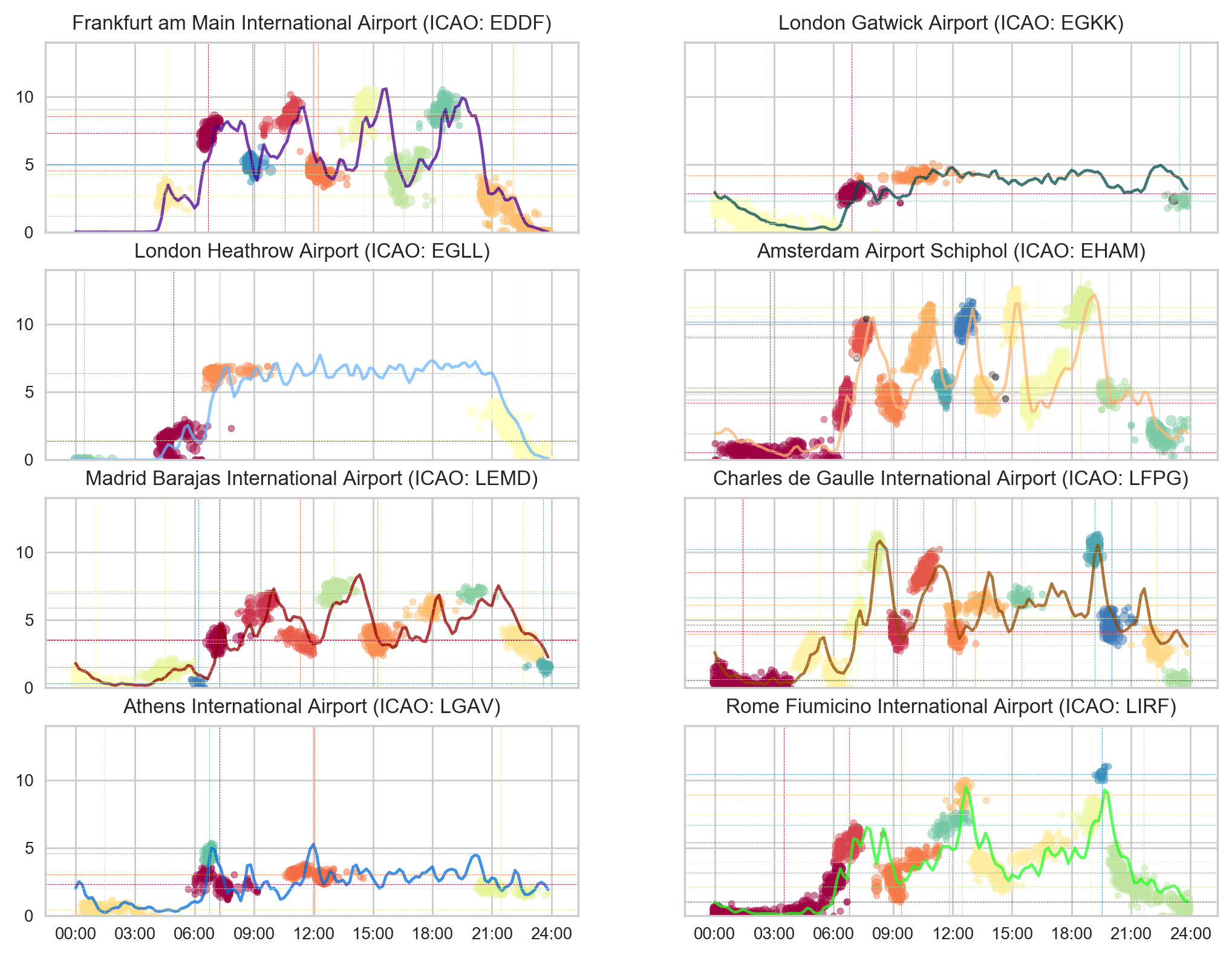}
            \caption{Data-driven modeling of non-homogeneous Poisson process.
            The \PELT{} algorithms returns a sequence of couples \((t_k,\lambda_{t_k})\), where \(\lambda_{t_k}\) is the intensity of the Poisson process in the interval \([t_k, t_{k+1})\). The couples are successively clustered via \DBSCAN{} (outliers are not displayed). The clusters in the plane \((t,\lambda)\) are shown superimposed to the daily average arrivals of each airport.}\label{fig:poisson_segmentation}
        \end{figure}

        Figure~\ref{fig:poisson_segmentation} shows the \DBSCAN{} clustering of changepoints identified by \PELT{} and the daily average rate of arrivals at 40 NM per 10-minute  intervals.
        For each cluster, thin dashed lines mark the average time of the day and the corresponding average Poisson intensity, whose values are reported by Table~\ref{tab:poisson_segmentation}.
        In view of the 24-hour periodicity of the demand highlighted in Section~\ref{sec:serial_corr}, we define our data-driven Poisson model by a periodic step-function \(\lambda^\ast(t)\) that takes on value \(\hat{\lambda}_i\) for \(\hat{t}_i \leq t < \hat{t}_{i+1}\).

        Table~\ref{tab:poisson_segmentation} gives evidence that
        \begin{enumerate}
            \item morning and afternoon regimes are substantially in line with the fitted values from Table~\ref{tab:fitted}, since \(\lambda \approx 60 \times \text{mean}^{-1}\) in the approximation of exponential interarrivals;
            \item these regimes are reflecting the declared capacity of the airports: for \airp{egll}, \(\lambda_k = 0.67\) aircraft/min corresponds to 40 aircraft/hour, which is a good proxy of the hourly landing rate; this remark validates the analysis carried-out so far because most of the airports considered work very close to their declared capacity.
        \end{enumerate}

        \begin{table}[tbp]
            \centering
            \caption{Non-homogeneous Poisson process derived by \PELT{} and \DBSCAN{} algorithms. The table reports the centroids of the clusters identified by \DBSCAN{} and shown in Figure~\ref{fig:poisson_segmentation}. Times are local, \(\lambda\) is measured in aircraft/min.}\label{tab:poisson_segmentation}
            \begin{tabular}{ccr|ccr}
\toprule
\textsc{icao} & time & \(\lambda\) &
\textsc{icao} & time & \(\lambda\)\\
\midrule
\airp{eddf} & 04:32 \textsc{utc}+02 &  0.2657 &
     \airp{lfpg} & 01:24 \textsc{utc}+02 &  0.0530 \\
     & 06:41 \textsc{utc}+02 &  0.7325 &
             & 05:18 \textsc{utc}+02 &  0.1852 \\
     & 08:55 \textsc{utc}+02 &  0.4991 &
             & 07:08 \textsc{utc}+02 &  0.5127 \\
     & 10:33 \textsc{utc}+02 &  0.8550 &
             & 08:06 \textsc{utc}+02 &  1.0003 \\
     & 12:14 \textsc{utc}+02 &  0.4530 &
             & 09:11 \textsc{utc}+02 &  0.4170 \\
     & 14:31 \textsc{utc}+02 &  0.8757 &
             & 10:32 \textsc{utc}+02 &  0.8515 \\
     & 16:31 \textsc{utc}+02 &  0.4270 &
             & 12:11 \textsc{utc}+02 &  0.3950 \\
     & 18:29 \textsc{utc}+02 &  0.9034 &
             & 13:06 \textsc{utc}+02 &  0.6116 \\
     & 22:04 \textsc{utc}+02 &  0.1182 &
             & 15:29 \textsc{utc}+02 &  0.6669 \\
\airp{egkk} & 02:40 \textsc{utc}+01 &  0.0671 &
         & 19:09 \textsc{utc}+02 &  1.0227 \\
     & 06:54 \textsc{utc}+01 &  0.2858 &
             & 20:01 \textsc{utc}+02 &  0.4632 \\
     & 10:10 \textsc{utc}+01 &  0.4195 &
             & 22:16 \textsc{utc}+02 &  0.3120 \\
     & 23:26 \textsc{utc}+01 &  0.2328 &
             & 23:21 \textsc{utc}+02 &  0.0631 \\
\airp{egll} & 00:25 \textsc{utc}+01 &  0.0000 &
     \airp{lgav} & 01:27 \textsc{utc}+03 &  0.0445 \\
     & 04:56 \textsc{utc}+01 &  0.1409 &
             & 06:45 \textsc{utc}+03 &  0.4592 \\
     & 07:16 \textsc{utc}+01 &  0.6410 &
             & 07:16 \textsc{utc}+03 &  0.2315 \\
     & 22:24 \textsc{utc}+01 &  0.1424 &
             & 12:03 \textsc{utc}+03 &  0.3028 \\
\airp{eham} & 02:47 \textsc{utc}+02 &  0.0537 &
         & 21:27 \textsc{utc}+03 &  0.1993 \\
     & 06:30 \textsc{utc}+02 &  0.4222 &
     \airp{lirf} & 03:29 \textsc{utc}+02 &  0.1019 \\
     & 07:25 \textsc{utc}+02 &  0.9062 &
             & 06:47 \textsc{utc}+02 &  0.5395 \\
     & 08:53 \textsc{utc}+02 &  0.4416 &
             & 09:24 \textsc{utc}+02 &  0.3155 \\
     & 10:28 \textsc{utc}+02 &  0.9006 &
             & 11:49 \textsc{utc}+02 &  0.6698 \\
     & 11:30 \textsc{utc}+02 &  0.5334 &
             & 12:30 \textsc{utc}+02 &  0.8899 \\
     & 12:38 \textsc{utc}+02 &  1.0207 &
             & 14:48 \textsc{utc}+02 &  0.3965 \\
     & 13:35 \textsc{utc}+02 &  0.4785 &
             & 18:58 \textsc{utc}+02 &  0.7465 \\
     & 15:00 \textsc{utc}+02 &  1.0653 &
             & 19:31 \textsc{utc}+02 &  1.0422 \\
     & 16:23 \textsc{utc}+02 &  0.5189 &
             & 21:37 \textsc{utc}+02 &  0.2130 \\
     & 18:26 \textsc{utc}+02 &  1.1255 \\
     & 19:52 \textsc{utc}+02 &  0.4881 \\
     & 22:26 \textsc{utc}+02 &  0.1942 \\
\airp{lemd} & 01:00 \textsc{utc}+02 &  0.0386 \\
     & 04:30 \textsc{utc}+02 &  0.1240 \\
     & 06:12 \textsc{utc}+02 &  0.0293 \\
     & 07:14 \textsc{utc}+02 &  0.3511 \\
     & 09:19 \textsc{utc}+02 &  0.5870 \\
     & 11:18 \textsc{utc}+02 &  0.3470 \\
     & 13:02 \textsc{utc}+02 &  0.7084 \\
     & 15:14 \textsc{utc}+02 &  0.3574 \\
     & 17:56 \textsc{utc}+02 &  0.5905 \\
     & 20:01 \textsc{utc}+02 &  0.6972 \\
     & 22:33 \textsc{utc}+02 &  0.3309 \\
     & 23:35 \textsc{utc}+02 &  0.1493 \\
\bottomrule
\end{tabular}

        \end{table}

        \subsection{Data-driven PSRA and comparison with Poisson}\label{sub:data_driven_psra}

\begin{figure}
            \includegraphics[width=\textwidth]{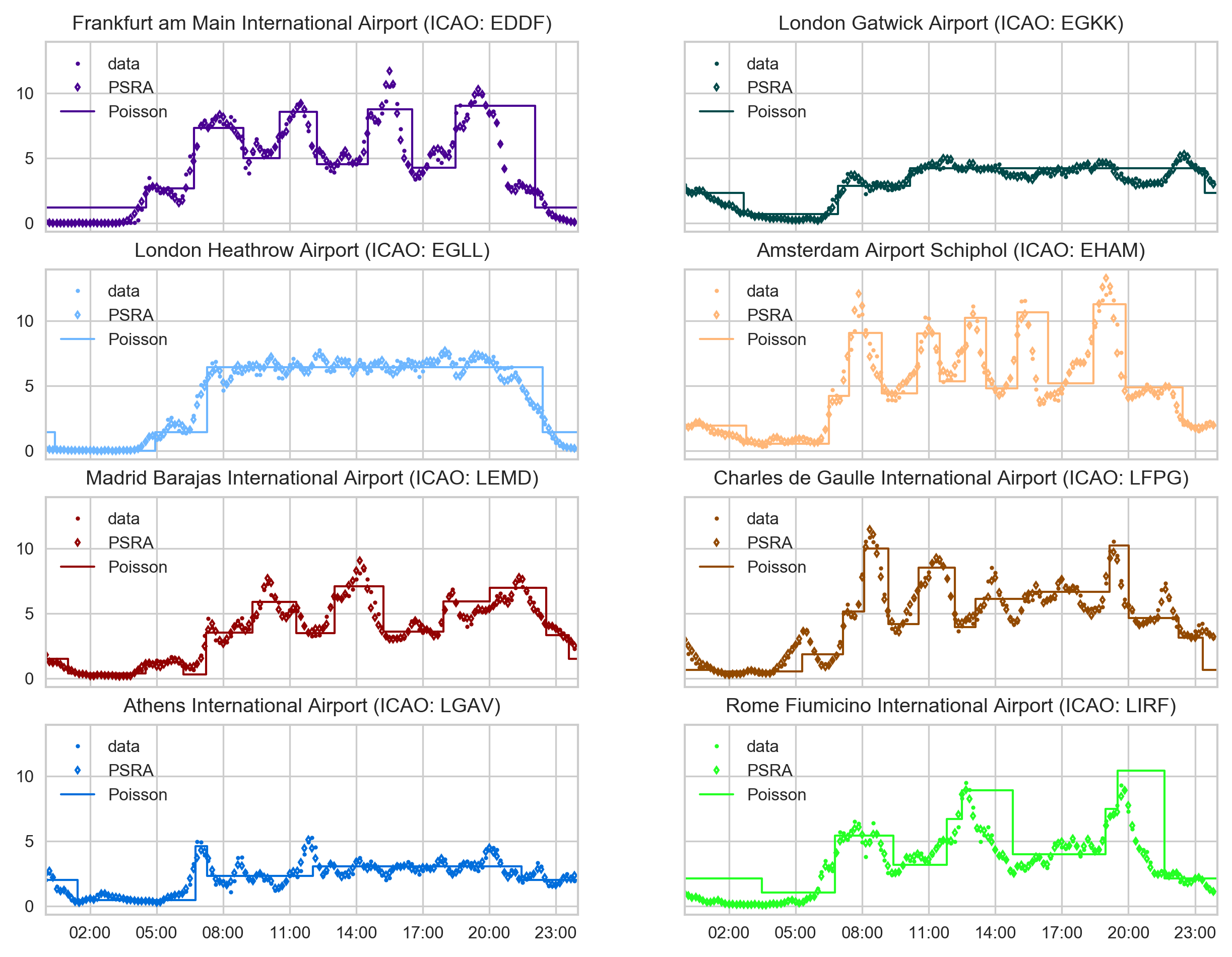}
            \caption{Comparison of the daily average arrival rate per interval of 10 minutes according to data, simulations of~\eqref{eq:psra-like}, and Poisson process.
            The Poisson process is defined by Table~\ref{tab:poisson_segmentation}.}\label{fig:mean_simul_arrivals}
        \end{figure}

        Figure~\ref{fig:mean_simul_arrivals} depicts the daily average demand obtained by averaging the arrivals \ac{TS}, the average daily demand  predicted by the data-driven Poisson process, and the average demand obtained by simulating the \ac{PSRA}~\eqref{eq:psra-like}.
        The latter is capable of reproducing the average demand of each airport with very good precision.
        This is not the case with the Poisson process defined by Table~\ref{tab:poisson_segmentation}, because the average number of arrivals is constant between \(\hat{t}_{i}\) and \(\hat{t}_{i+1}\).

\begin{figure}
            \includegraphics[width=\textwidth]{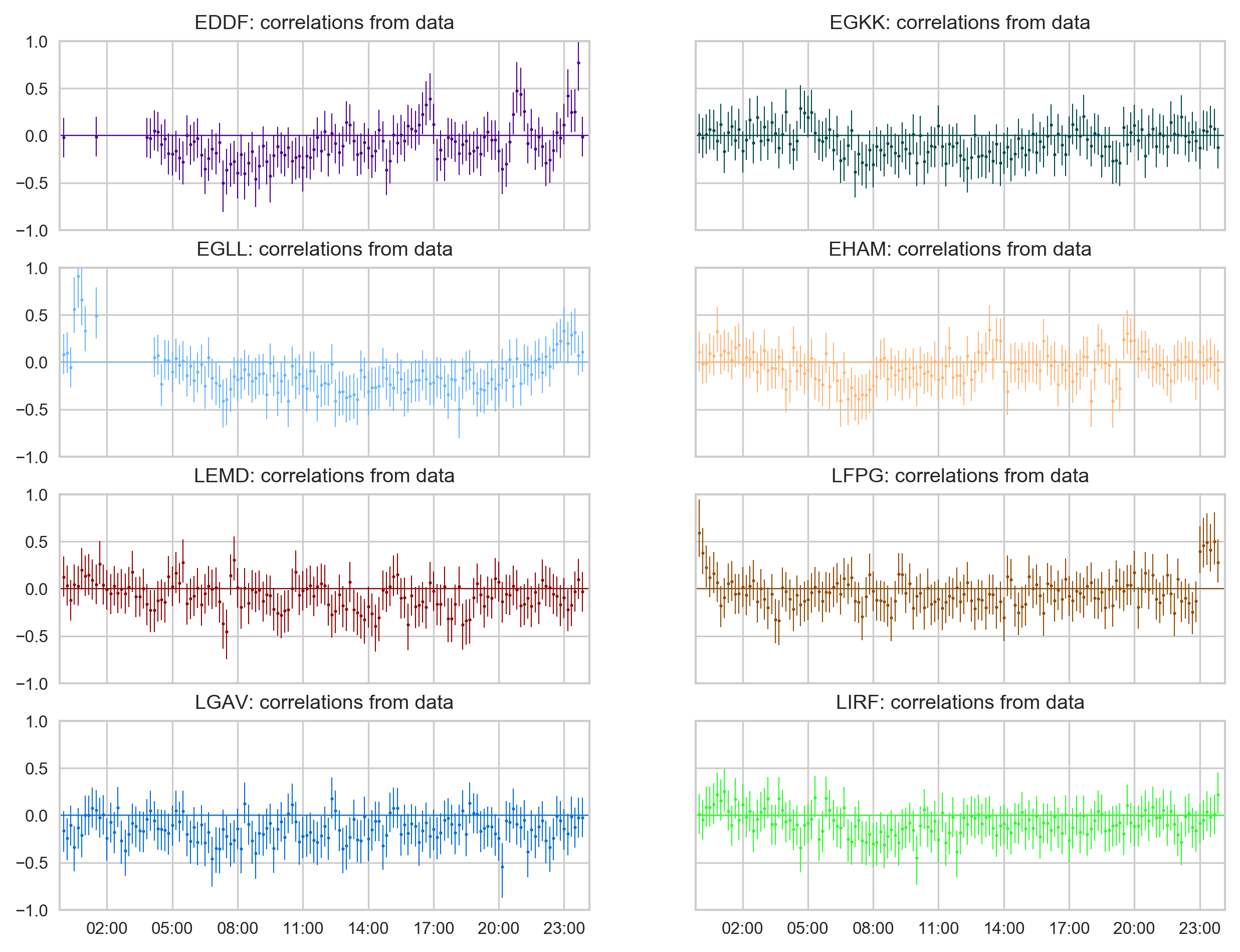}
            \caption{Correlations from \(t^{M3}_i\) data. Error bars show 95\% confidence interval for Pearson's \(\rho\).}\label{fig:correlations_true}
        \end{figure}

\begin{figure}
            \includegraphics[width=\textwidth]{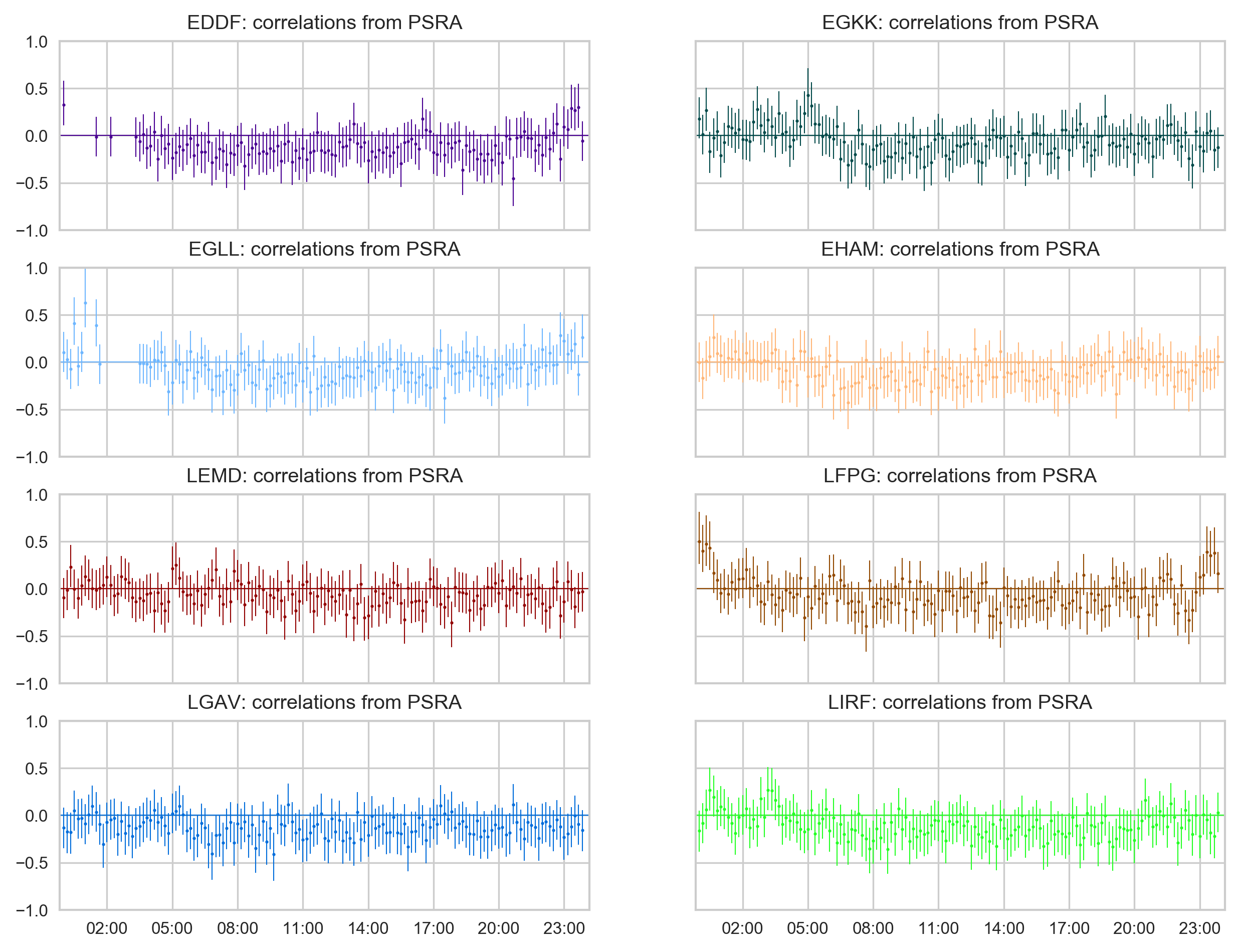}
            \caption{Correlations from simulation of Model~\eqref{eq:psra-like}. Error bars show 95\% confidence interval for Pearson's \(\rho\).}\label{fig:correlations_psra}
        \end{figure}

\begin{figure}
            \includegraphics[width=\textwidth]{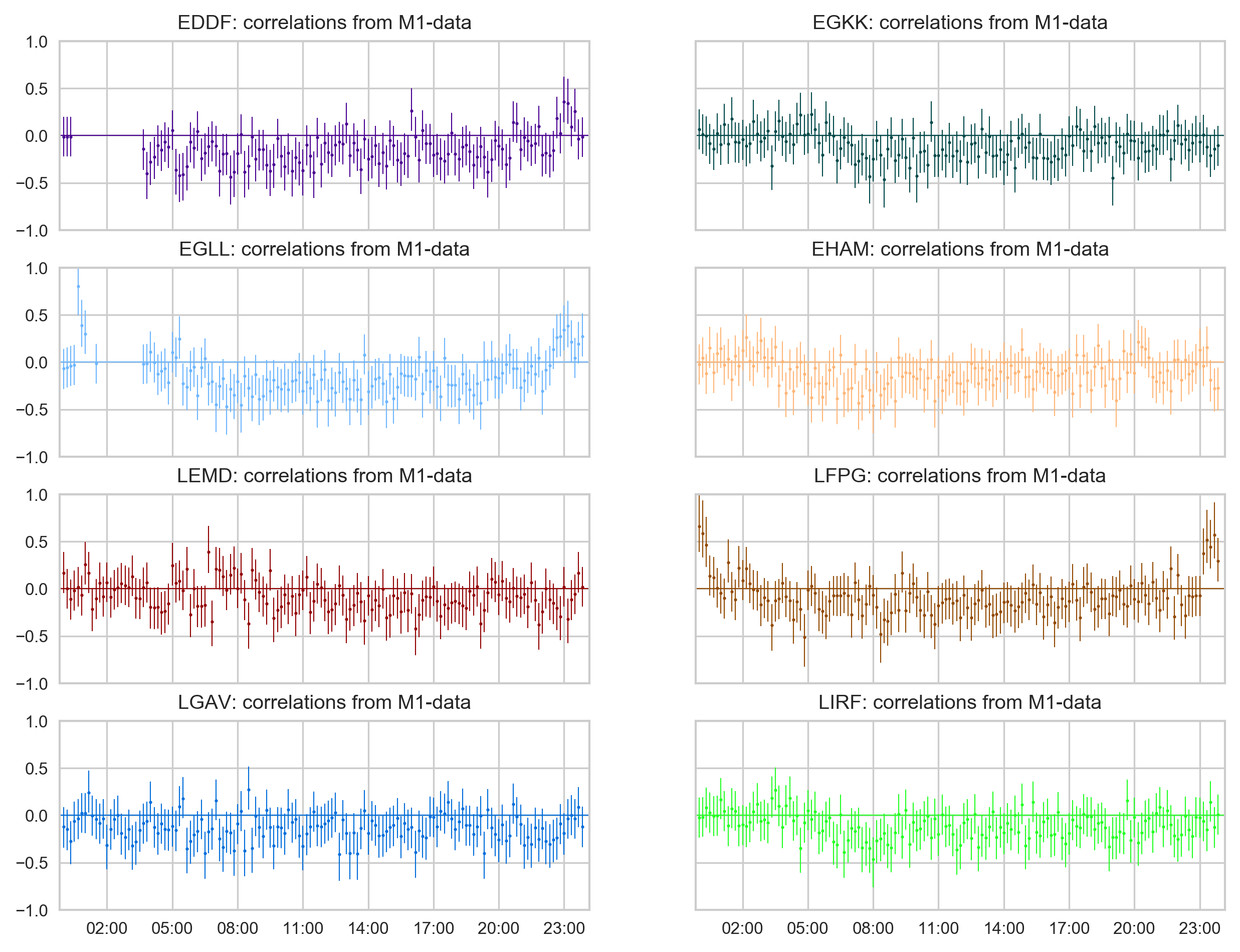}
            \caption{Correlations from \(t^{M1}_i\) data. Error bars show 95\% confidence interval for Pearson's \(\rho\).}\label{fig:correlations_m1}
        \end{figure}

        Figures~\ref{fig:correlations_true}-\ref{fig:correlations_m1} show the Pearson's correlation \(\rho_{t_i, t_{i+1}}\) between the number of arrivals in the intervals \([t_i, t_{i+1})\) and \([t_{i+1}, t_{i+2})\).
        The value of \(\rho_{t_i, t_{i+1}}\) is computed from \(t^{M3}_i\) data (Fig.~\ref{fig:correlations_true}), from simulation of Model~\eqref{eq:psra-like} (Fig.~\ref{fig:correlations_psra}), and from \(t^{M1}_i\) data (Fig.~\ref{fig:correlations_m1}).
        The value of \(\rho_{t_i, t_{i+1}}\) is generally statistically different from zero at a 5\% significance level, and the point estimates are generally negative.
        This finding matches with lag-1 autocorrelations reported by Table~\ref{tab:lag01}.
        Also, the sign of the correlations agrees with the finite capacity of both terminal airspace-sectors and airport.
        In fact, if the arrivals were mostly uncorrelated or even positively correlated, the capacity of airports and/or  sectors might be exceeded by random fluctuations in the inbound flow.
        Instead, the airspace finite capacity imposes a structure in the sequence of arrivals.
        This structure can be seen in Figure~\ref{fig:correlations_m1}, where \(\rho_{t_i, t_{i+1}}\) is computed from the \(M1\) flight plans, \ie{} the \ac{TS} of last-agreed arrivals at 40 NM\@.
        In this respect, the \ac{PSRA}-like model~\eqref{eq:psra-like} is very close to capturing the actual characteristics of the inbound flow because, by superimposing \ac{IID} random fluctuations to the last-agreed arrivals, model~\eqref{eq:psra-like} produces a stream of correlated arrivals that inherits the \emph{original} correlation structure of the data.


    \section{Analysis and Conclusions}\label{sec:conclusions}

    The paper has analyzed a dataset of inbound traffic data at ten important European airports.
    Having in mind the description of airport congestion through a queuing system, we have modeled the inbound stream of aircraft in a data-driven fashion.
    For simplicity, we have considered the entrance time of an idealized cylinder of radius 40 NM centered at the arrival airport.
    The analytical tools used and the methodological approach highlighted above would not be affected if that time was to be replaced by a the first entrance-time in a geographically-specified airspace volume.

    Analysis of the interarrival times at 40 NM show a fair accordance of the latter with a Weibull distribution; this is already known in the literature for \emph{projected} interarrivals at some major US airports~\citep{willemain2004statistical}.
    The shape parameter of the fitted Weibull is sufficiently close to unity to seemingly justify the \emph{classical} hypothesis of Poisson arrivals.
    The null hypothesis of Weibull-distributed \ac{IID} interarrivals can be often rejected based on the Kolmogorov-Smirnov goodness-of-fit test, but this observation requires some additional comments.
    Due to the large sample size, the test is extremely powerful (especially at the most congested airports) and thereby capable of detecting even very small differences between the empirical and the theoretical distribution.
    The average demand plotted in Figure~\ref{fig:AvgArrivals} suggests that the poor fit might be related to the wavy variation of the demand.
    On the one hand, the QQ-plots in Figures~\ref{fig:qqplot5-8}-\ref{fig:qqplot5-17} show that the best fit is achieved by airports where the demand is stable over time, \eg{} \airp{egkk} or \airp{egll}; on the other, the selected time windows often span moments of lower and higher demand, so that a better fit might be probably obtained with a mixture of Weibulls.
    The complicacy of the latter model lies beyond the scope of the paper, yet the previous remark and the variability of the fitted Weibull's parameter throughout the day suggest that the underlying arrival process can be possibly described by a non-homogeneous Poisson process.
    The findings above seem also to give indication that the parameter of a non-homogeneous Poisson process should vary quite fast in time so to match the variability of the demand.

    We have proposed an innovative, data-driven approach to the modeling of a non-homogeneous Poisson process by use of the \PELT{} (changepoint algorithm) and \DBSCAN{} (clustering).

    The Poisson arrivals defined this way were contrasted with a \ac{PSRA} point process, where the observed arrival time arises as the sum of the last-agreed arrival time and a random fluctuations.
    Instead of using a parametric distribution for the fluctuations, we have taken \ac{IID} samples from the empirical distribution of the delays.
    The model could  be easily improved
    by sampling the \(\xi_i\) from a stratified empirical distribution, that is, defining \(\xi = \xi(X)\), where \(X\) is a vector of explanatory variables, like
    time of the day, en-route whether conditions, departure airport, etc.
    Provided that a sufficiently large sample is available, stratified sampling is likely to define a more precise model.

    The \ac{PSRA} model has an extremely simple, non-parametric formulation, yet it succeeds in reproducing the typical arrivals pattern.
    It outperforms the Poisson process with respect to the capacity of reproducing the average number of arrivals in a given time interval.
    The level of accuracy of the Poisson process could be increased by modeling the intensity of the process on a finer time scale, i.e., by estimating \(\lambda^\ast(t)\) in small prescribed intervals --it is well known that the \ac{MLE} of the parameter \(\lambda\) is the sample mean.
    Thus, a Poisson process with \(\lambda(t)\) varying every 10 minutes would be a model that exactly reproduces the daily average demand.
    In other words, a Poisson process could reproduce the typical demand more accurately if the intensity function \(\lambda(t)\) were \emph{forced} to vary on a fine time-scale.
    Nevertheless, this model would have 144 parameters and it would still fail to capture the correlation structure of the arrival data.
    This is because a non-homogeneous Poisson process has independent increments regardless of the functional form of the intensity \(\lambda(t)\).
    In contrast, the \ac{PSRA} model naturally captures the original correlations of the arrivals and has the advantage of being non-parametric --or requiring a low number of parameters if defined like in~\citet{caccavale2014model,guadagni2011queueing}.

    This paper has presented many arguments to adopt a \ac{PSRA} point process for modeling the aircraft inbound flow.
    Although interesting results have been recently presented in~\citet{lancia2013advances} for the case of exponential \(\xi_i\)'s, the current main limitation to the use of the \ac{PSRA} family is the great difficulty of any analytical treatment as soon as the distribution of \(\xi\) is non-trivial.
    Nevertheless, a non-homogeneous Poisson process would likely be affected by a similar problem if \(\lambda(t)\) varied too rapidly, making the equilibrium analysis of any regime of no practical use.
    The final outcome of the paper is thus a general recommendation of preferring, whenever possible, the adoption of \ac{PSRA} for the modelling and simulation-based analyses of the inbound traffic over any process in the family of Poisson arrivals.

    \appendix
    \section{Appendix: PELT with MBIC and AIC penalty}\label{sec:appendix}

\begin{figure}
        \centering
        \includegraphics[width=\textwidth]{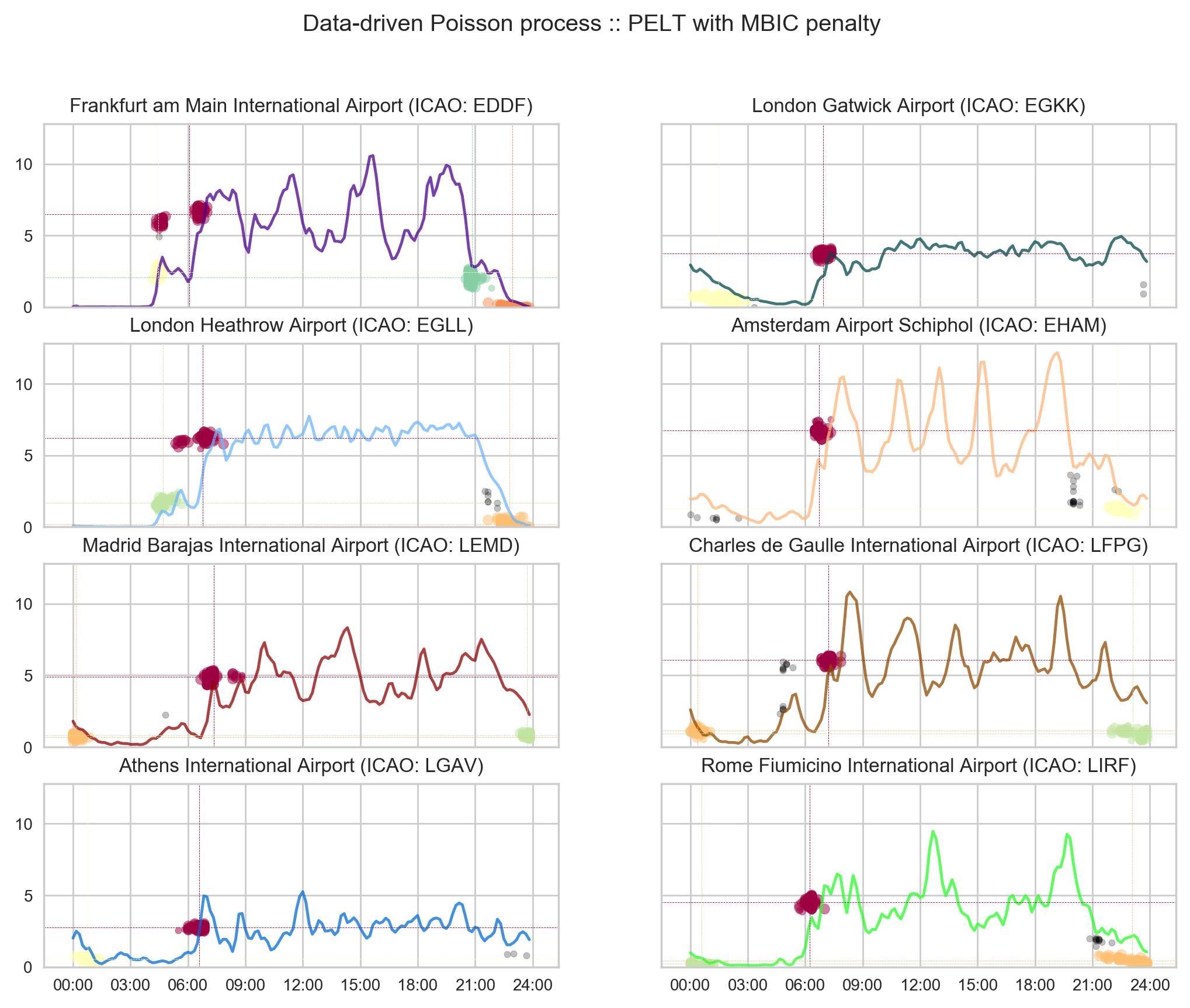}
        \caption{Data-driven modeling of non-homogeneous Poisson process; \ac{MBIC} penalty. \DBSCAN{} outliers are drawn in black.}\label{fig:DD_MBIC}
    \end{figure}

    This appendix offers a sensitivity analysis of the \PELT{} algorithm with respect to the result of Algorithm~\ref{Alg:POISSON}.
    Figure~\ref{fig:DD_MBIC} shows the changepoint identified by \PELT{} and the resulting clustering via \DBSCAN{} when \PELT{} uses the default \ac{MBIC} penalty.
    In this case, the changepoints returned by \PELT{} are so well-concentrated that the clustering step is barely needed.
    Accordingly, the number of outliers returned by \DBSCAN{} (shown in black) is extremely limited.
    Identified clusters are limited in number and typically located early in the morning or late in the evening.
    Thus, the Poisson intensities returned by Algorithm~\ref{Alg:POISSON} have a very straightforward interpretation as day and night regimes, possibly with an intermediate value like \airp{eddf} and \airp{egll}.

\begin{figure}
        \centering
        \includegraphics[width=\textwidth]{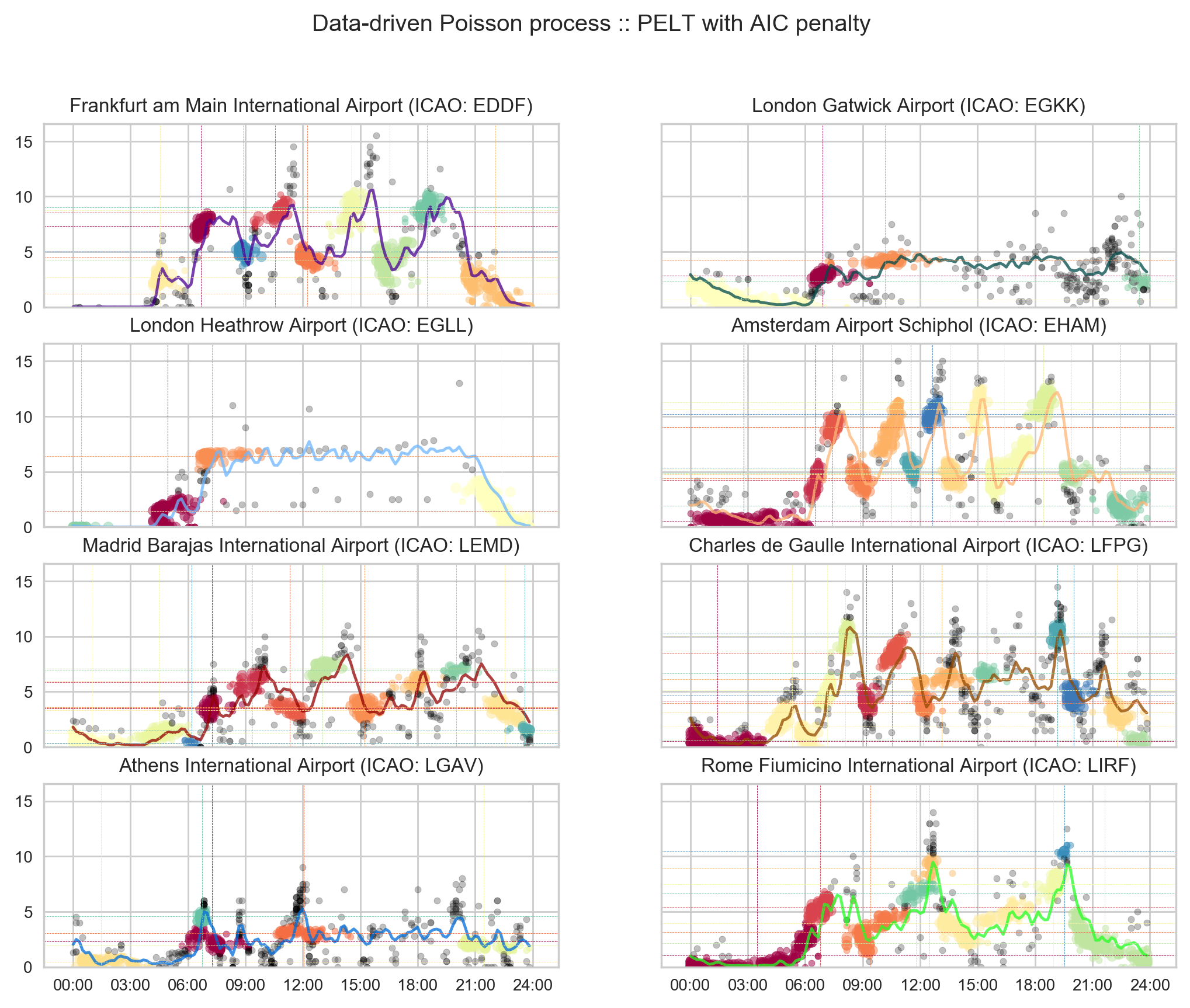}
        \caption{Data-driven modeling of non-homogeneous Poisson process; \ac{AIC} penalty. \DBSCAN{} outliers are drawn in black.}\label{fig:DD_AIC}
    \end{figure}

    Figure~\ref{fig:DD_AIC} shows the changepoint identified by \PELT{} and the resulting clustering via \DBSCAN{} when \PELT{} uses the \ac{AIC} penalty.
    In this setting, \PELT{} is much more sensible and detects regime changes in correspondence of maxima and minima of the average demand.
    The resulting description of the arrivals stream in terms of a Poisson process is hence richer and follows more closely the average demand.
    However, the increased sensibility in the changepoint detection comes at the price of a \emph{noisy} distribution of changepoints in the \((t,\lambda)\) plane, which might be diffcult to reconstruct via \DBSCAN{} (see \airp{egkk}).

    \addtocontents{toc}{\protect\vspace{\beforebibskip}}
    \addcontentsline{toc}{section}{\refname}
    \bibliographystyle{plainnat}
    \bibliography{ll}

\end{document}